\documentstyle[11pt,epsfig]{article}
\def\arraystretch{1.2}
\begin{document}
\begin{titlepage}
\hfill SNUTP97/032
\vfill
\begin{center}
{\Large\bf Strange Form Factors of Baryons}\\
\vskip 8.mm
{Soon-Tae Hong$^{a}$, Byung-Yoon Park$^{b}$ and Dong-Pil Min$^{a}$}
\vskip 0.4cm
{$^{a}$Department of Physics and Center for Theoretical Physics,}\par
{Seoul National University,}
{Seoul 151-742, Korea}
\vskip 0.3cm
{$^{b}$Department of Physics,}\par
{Chungnam National University,}
{Daejeon 305-764, Korea}
\vskip 1.0cm
{\today}
\vskip 1.5cm
\vfill
{\bf ABSTRACT}
\begin{quotation}
The strange magnetic form factor of proton is calculated in a model 
independent way to confirm the recent experimental result of the SAMPLE 
Collaboration.  We consider a set of six inertia parameters to realize the 
magnetic moments of the baryon octet.  We show that the strange form factor of 
proton is a positive quantity, i.e. +0.37 n.m..  Its positiveness is analyzed 
in terms of the vacuum fluctuation coupled to the vector current along the 
strangeness direction.  
\vskip 1.0cm
\noindent
PACS; 21.60; 13.40.G\\
\noindent
Keywords; strange form factors; chiral bag
\end{quotation}
\end{center}
\end{titlepage}
 
Triggered by the EMC result\cite{emc}, there have been significant discussions 
concerning the possibility of sizable strange quark matrix elements in the 
nucleon.  Quite recently, the SAMPLE experiment\cite{sample} reported the 
proton's neutral weak magnetic form factor, which has been suggested by the 
neutral weak magnetic moment measurement through parity violating electron 
scattering\cite{mck89}.  In fact, if the strange quark content in the nucleon 
is substantial then kaon condensation can be induced at a matter density lower 
than that of chiral phase transition\cite{kaon1,kaon2} affecting the scenarios
for relativistic heavy-ion reactions\cite{heavy}, neutron star 
cooling\cite{neu} and so on.

On the other hand, McKeown\cite{mck} has shown that the strange form factor of 
proton should be positive by using the conjecture that the up-quark effects are 
generally dominant in the flavor dependence of the nucleon properties.  
This result is contrary to the negative values of the proton 
strange form factor which result from most of the model 
calculations\cite{model1,model2,model3,model4,model5,model6}
except that of Hong and Park (HP)\cite{hp} based on the 
SU(3) chiral bag model (CBM)\cite{cbm}.  Recently Meissner {\it et al} 
investigated the strange form factors by taking isoscalar and isovector meson 
poles into account\cite{meissner}.  They found the strange form factor to be 
positive and quite ``OZI-resurrected", $\mu_s \approx 0.003$ n.m..

The motivation of this paper is to justify the prediction of Ref.\cite{hp} on 
the positive strange form factor of the proton by doing the adjustment of the
inertia parameters in a more systematic way.

Long ago, Adkins discussed the model independent parameterization of the 
magnetic moments of baryons governed by the group structure, in which the 
flavor symmetry breaking and the 
$1/N_c$ correction are taken into account of the SU(3) Skyrme model 
(SM)\cite{adkins}.  The same line was traced in HP for the CBM to show that 
there are also six parameters for the magnetic moments of baryon octet. As the 
details of  the calculation with CBM already are presented in HP, we here 
mention only the general scheme of it to discuss the model-independent behavior of the magnetic moments.

In the CBM, a baryon is described by two phases : the mesonic phase of soliton 
configuration is surrounding the quark phase where the freely moving quarks are
confined through the chirally symmetric boundary condition.  In addition to the
``chirally symmetric" Lagrangian, incorporated are various ``symmetry 
breaking" terms into the Lagrangian: (1) Breaking of the SU(3)$_L \times$
SU(3)$_R$ chiral symmetry into SU(3)$_V$ for the nonvanishing quark masses. At 
this level all the quark masses and consequently the masses of the meson octet
are chosen degenerate. The mass matrix $\bar{M}$ is given by $\bar{M} = \bar{m}
\mbox{diag}(1,1,1)$ with $\bar{m}=\frac12(m_u + m_d)$.  (2) Breaking of the 
flavor SU(3) symmetry into SU(2)$\times$U(1) due to the heavier strange quark 
than the other two light quarks and due to the difference in the kaon and pion 
decay constants $f_K\neq f_\pi$.  The isospin symmetry 
breaking has not been incorporated.  These symmetry breaking terms can be 
treated as perturbations. 

The meson part of the solution is described  
by a classical static field configuration 
$U_0(\vec{r}) = \exp(i\sum_{i=1}^3 \lambda_i \hat{r}_i \theta(r))$ 
and the quark part is described by 
a Fock state $|H\rangle_0$ with $N_c$ valence 
quarks and completely filled negative energy sea for the quarks 
confined inside the bag.  
The zero modes associated with invariance of the solution 
under the arbitrary rotations in SU(3) flavor space 
can be canonically quantized 
by introducing proper collective coordinates $A(t) \in $SU(3). 
Such a process leads us to the Hamiltonian for the baryon states as 
\begin{equation}
\hat{H} = \hat{H}_{0} + \hat{H}_{SB},
\label{H}\end{equation}
where
$$\hat{H}_0 =  M_0
	    + \frac12 \left(\frac{1}{{\cal I}_1}
			  - \frac{1}{{\cal I}_2}\right) \hat{J}^2 
	    + \frac{1}{2{\cal I}_2} (\hat{C}^2_2
		       - \textstyle\frac34 \hat{Y}_R^2), 
\eqno(\mbox{\ref{H}a})$$
and
$$\hat{H}_{SB} = m [1-\hat{D}^{8}_{88}(A)]
	       - m_1 \hat{Y}.
\eqno(\mbox{\ref{H}b})$$
Here, the operators associated with the SU(3) collective 
coordinate quantization will be distinguished by putting a caret : 
$\hat{C}^2_2$ is the quadratic Casimir operator for flavor SU(3), 
$\hat{J}_i(i=1,2,3)$ the spin operator, 
$\hat{Y}_R$ the ``right" hypercharge operator, 
$D^{8}_{ab}(A)$ the adjoint representation of SU(3), 
and $\hat{Y}$ the hypercharge operator.  
The other quantities are the inertia parameters : 
$M_0$ is the mass of the soliton solution,
${\cal I}_1$ and ${\cal I}_2$, respectively, are its moments of 
inertia with respect to the collective rotation in the 
nonstrangeness and strangeness directions, 
and $m$ and $m_1$ are inertia parameters associated with the flavor symmetry 
breaking (FSB).  (See Ref.\cite{cbm} for their explicit form.) 
Except these model dependent inertia parameters, 
the Hamiltonian of Eq.(\ref{H}) is general to  
all the soliton models for baryons 
based on the chiral Lagrangian.

As far as the symmetric part of the Hamiltonian $H_0$ is concerned, 
the wavefunction of the baryon with isospin ($I$, $I_3$),  
spin ($J$, $J_3$) and the hypercharge $Y$ is given by the 
Wigner $D$-function, 
\begin{equation}
\Phi_B^{(\lambda)}(A) = \sqrt{\lambda} 
\langle I, I_3, Y|D^{(\lambda)}(A) |J,-J_3,+1\rangle,
\end{equation}
where $D^{(\lambda)}(A)$ is the matrix element of 
the SU(3) irreducible representation (IR) of dimension $\lambda$ 
acting on the basis $\langle I,I_3,Y|$ and $|J,-J_3,Y_R\rangle$. 
Due to the Wess-Zumino constraint, 
only the states obeying $Y_R=1$ are allowed.
When the symmetry breaking Hamiltonian $\hat{H}_{SB}$
is treated as a perturbation, 
up to first order the wavefunctions of the baryon octet are 
modified as 
\begin{equation}
\Phi_B(A) = \Phi_B^{(8)}(A)
	  - C_{(\underline{10})} \Phi_B^{(\underline{10})}(A) 
	  - C_{(27)} \Phi_B^{(27)}(A),  
\label{wf}\end{equation}
where $\Phi_B^{(\underline{10})}(A)$ and $\Phi_B^{(27)}(A)$ have the 
same quantum numbers of the corresponding baryon  belonging to 
different IR.  
Such a ``representation mixing" is caused 
by the $\hat{D}^{8}_{88}(A)$ in $\hat{H}_{SB}$. 

The baryon magnetic moments can be calculated by taking the 
expectation values of the corresponding operator with respect 
to the baryon wavefunctions (\ref{wf}). 
We first derive the flavor singlet current $V^{(0)}_\mu$ 
and flavor octet vector current $V^{(a)}_\mu(a=1,2,\cdots,8)$ 
from the given CBM Lagrangian.
As for a baryon number current on the mesonic phase, 
we should  include that of the topological winding number.  
In terms of these current, the electromagnetic (em) 
currents $J^{em}_{\mu}$ for example  
can be easily constructed as 
$J_\mu^{em} = V^{(3)}_\mu + \frac{1}{\sqrt3} V^{(8)}_\mu$. 
The currents define associated magnetic moments as
\begin{equation}
\vec{\mu}^{(0,a)}=\mbox{$\frac{1}{2}$} \int{\rm d}^{3}r
(\vec{r} \times \vec{V}^{(0,a)}).
\end{equation}
Following the standard procedures in 
the SU(3) collective coordinate quantization scheme, 
one can obtain the magnetic moment operators in terms of the 
collective variables as
\begin{equation}
\begin{array}{lcl}
   \hat{\mu}^{i(0)}  
      &=&  {\cal M} \hat{J}^i, \\
   \hat{\mu}^{i(a)} 
      &=& \displaystyle 
      - {\cal N}\hat{D}_{ai}^{8}
      - {\cal N}^{\prime} \sum_{p,q=4}^7 d_{ipq} 
	      \hat{D}_{ap}^{8}\hat{T}_{q}^{R}
      + \frac{N_c}{2\sqrt3}{\cal M} \hat{D}_{a8}^{8}\hat{J}_{i} 
      \\
      & & \displaystyle 
      - {\cal P} \hat{D}_{ai}^{8} (1-\hat{D}_{88}^{8})
      + \frac{\sqrt3}{2}{\cal Q} \sum_{p,q=4}^7 
	    d_{ipq} \hat{D}_{ap}^{8} \hat{D}_{8q}^{8}
\end{array}
\label{muop}\end{equation}
where ${\cal M}$, ${\cal N}$, ${\cal N}^{\prime}$, ${\cal P}$ and ${\cal Q}$
are new inertia parameters depending on the soliton solution.  
(See Refs.\cite{hp,pmr} for the details.) 
$\hat{T}^R_p (p=4,\cdots,7)$ is the ``right" SU(3) operator appearing 
in the SU(3) collective coordinate quantization processes. 
The other operators such as $\hat{T}^R_i(i=1,2,3)$ and $\hat{T}_8^R$ 
have been replaced by familiar spin operator $\hat{J}_i$ and the  
right hypercharge operator $Y^R$ mentioned before :  
$\hat{J}_i = - \hat{T}^R_i$ and $\hat{Y}^R = 2\sqrt{\frac13} \hat{T}^R_8$. 
The last two terms of $\hat{\mu}^{i(a)}$ 
come from the symmetry breaking Lagrangian 
for the difference in the meson decay constants. 
In the practical calculations of the expectation values, it is helpful 
to rewrite the last two expressions in terms of a linear combination 
of the single $\hat{D}$ operators, for example, as
\begin{equation}
\begin{array}{rcl}
\hat{\mu}^{i(3)}_{SB}
   &=&{\cal P}(-\frac{4}{5} \hat{D}_{3i}^{8}
	       +\frac{1}{4}( \hat{D}_{3i}^{10}
			   + \hat{D}_{3i}^{\underline{10}})
	       +\frac{3}{10} \hat{D}_{3i}^{27})
      +{\cal Q}(\frac{3}{10} \hat{D}_{3i}^{8}
		- \frac{3}{10} \hat{D}_{3i}^{27}), \\
\hat{\mu}^{i(8)}_{SB}
   &=&{\cal P}(-\frac{6}{5} \hat{D}_{8i}^{8}
	       +\frac{9}{20} \hat{D}_{8i}^{27})
     +{\cal Q}(-\frac{3}{10} \hat{D}_{8i}^{8}
	       -\frac{9}{20} \hat{D}_{8i}^{27}).
\end{array}
\end{equation}
Here, $D_{ab}^{\lambda}$ is the unitary IR of SU(3),  
which is nothing but the Wigner $D$-functions appearing in Eq.(\ref{wf}) 
with the corresponding 
quantum numbers $(Y;I,I_{3})_a$ and $(Y_{R};J,-J_{3})_b$.

By taking the expectation values of the operators (\ref{muop}) 
with the wavefunctions (\ref{wf}) for the baryon states,
one can obtain the electromagnetic moments in a form of
\begin{equation}
\mu_B = \mu_{0,B} + \delta \mu_{1,B} + \delta \mu_{2,B}
\end{equation}
up to the first order in perturbation. 
Here, the first order correction $\delta \mu_{1,B}$ comes from the 
explicit contribution of the FSB Lagrangian 
to the current and $\delta \mu_{2,B}$ is due to the representation 
mixing in the wavefunctions.  In Table 1, listed are the explicit expressions 
for the magnetic moments of the baryon octet.  Note that $\delta\mu_{2,B}$ 
(and the term proportional to ${\cal Q}$ in $\delta\mu_{1,B}$) 
satisfy ``V-spin symmetric" relations
\begin{equation}
\delta \mu_{2,p} = \delta \mu_{2,\Xi^-}, \hskip 1em
\delta \mu_{2,n} = \delta \mu_{2,\Sigma^-} \hskip 1em
\mbox{and} \hskip 1em
\delta \mu_{2, \Sigma^+} = \delta \mu_{2, \Xi^0}
( = \textstyle\frac12 \delta \mu_{2,p}),
\label{v-spin}\end{equation}
while $\mu_{0,B}$'s satisfy SU(3) symmetric
(or ``U-spin symmetric") ones
\begin{equation}
\mu_{0,p} = \mu_{0,\Sigma^+}, \hskip 1em
\mu_{0,n} = \mu_{0,\Xi^0}, \hskip 1em
\mu_{0,\Sigma^-} = \mu_{0, \Xi^-} \hskip 1em
\mbox{and} \hskip 1em
\mu_{0,\Lambda} = - \mu_{0, \Sigma^0}.
\label{u-spin}\end{equation}

At this point, it will be interesting to evaluate
the separate up, down and strange quark contributions
to the magnetic moments of the baryons.
They are associated with the electromagnetic current carried by 
the quark of flavor $f(=u,d,s)$,
which are obtained by multiplying the fractional charge
to the corresponding quark current $V^{(f)}_\mu$; that is, 
$J^{em(f)}_\mu = Q^{(f)}_{em} V^{(f)}_\mu$ with 
$Q^{(u)}_{em}=+\frac23$, $Q^{(d,s)}_{em}=-\frac13$
and 
\begin{equation}
\begin{array}{l}
V^{(u)}_\mu = V^{(0)}_\mu + V^{(3)}_\mu 
	    + \frac{1}{\sqrt3} V^{(8)}_\mu, \\
V^{(d)}_\mu = V^{(0)}_\mu - V^{(3)}_\mu 
	    + \frac{1}{\sqrt3} V^{(8)}_\mu, \\
V^{(s)}_\mu = V^{(0)}_\mu - \frac{2}{\sqrt3} V^{(8)}_\mu.
\end{array}
\label{v-uds}\end{equation}
Let $\mu^{(f)}_B$ denote the each quark contribution
to the magnetic moment, which can be obtained in the same way 
described above.  
The explicit expressions for $\mu^{(s)}_B$ 
are appended in Table 2.
Note that the baryons belonging to the same isospin 
multiplets have the same strange component. 
The rest up- and down-quark contributions
can be obtained with the help of    
$\mu_B = \mu^{(u)}_B + \mu^{(d)}_B + \mu^{(s)}_B$
and by using the isospin symmetry :
\begin{equation}  
\mu_{B}^{(d)} = \frac{Q^{(d)}_{em}}{Q^{(u)}_{em}} 
		\mu_{\tilde{B}}^{(u)}, 
\label{ud}
\end{equation}
where $\tilde{B}$ denotes for the isospin conjugate baryon to $B$.

The form factors are defined through 
\begin{equation}
{}_B \langle p+q|V^{(f)}_\mu |p\rangle_B = \bar{\psi}_B 
\left[ \gamma_\mu F^{(f)}_{1,B}(q^2) 
       +\frac{i}{2m_B} \sigma_{\mu\nu} q ^\nu F^{(f)}_{2,B}(q^2) 
\right] \psi_B, 
\end{equation}
where $\psi_B$ is the spinor for the baryons. 
In the limit of $q^2 \rightarrow 0$, $\mu^{(f)}_B$ are related to the form 
factors $F^{(f)}_{2,B}(q^2)$ discussed in the literature. 
The relation between the form factors $F^{(f)}_{i,B}(i=1,2)$ and  
the each component of the magnetic moments $\mu^{(f)}_B$ reads 
\begin{equation}
\mu^{(f)}_{B} = Q^{(f)}_{e.m} \left[
F^{(f)}_{1,B}(0) + F^{(f)}_{2,B}(0) \right].
\end{equation}
On the other hand, the form factor $F^{(f)}_{1,B}$
is just the number of  valence quarks with the flavor $f$  contained in the baryon,
which is trivially given by its quantum numbers as 
\begin{equation}
\textstyle
F^{(u)}_{1,B}(0) = 1 + I_3 + \frac12 Y, \hskip 2em
F^{(d)}_{1,B}(0) = 1 - I_3 + \frac12 Y, \hskip 2em
{\rm and} \hskip 2em 
F^{(s)}_{1,B}(0) = 1 - Y.
\end{equation}
Thus, the strange quark contributions to the electromagnetic moments 
can be transformed into
those for the $F^{(s)}_{2,B}$ form factors by 
\begin{equation}
\begin{array}{ll}
F^{(s)}_{2,N}(0) = -3\mu^{(s)}_N, &
F^{(s)}_{2,\Lambda}(0) = -3\mu^{(s)}_\Lambda-1, \\
F^{(s)}_{2,\Sigma}(0) = -3\mu^{(s)}_\Sigma-1, &
F^{(s)}_{2,\Xi}(0) = -3\mu^{(s)}_\Xi -2. 
\end{array}
\label{f}\end{equation}
Note that the strange magnetic moment of nucleon comes solely from the 
$ F^{(s)}_{2,N}(0)$.  Furthermore, the I-spin symmetric relation (\ref{ud}) can be expressed in a simpler form as 
\begin{equation}
F^{(u)}_{2,B}(0) = F^{(d)}_{2,\tilde{B}}(0).
\end{equation}

We summarize the expressions for the magnetic moments of baryons 
in Table 1  which are general to various
chiral soliton models of baryons.
Then, we optimize the inertia parameters to fit the data in the point of view that
the situation can be improved by fine-tuning the model,
for example, by including the other degrees of freedom and/or higher order terms
in the derivative on the chiral field. 
In Ref.\cite{hp}, this trial was done by varying roughly the inertia parameters 
around the CBM values to minimize the sum of the absolute
difference between the model predictions and the measured magnetic moments of 
the baryons\footnote{Ref.\cite{hp} has some erroneous numerical values in the
predictions of the baryon magnetic moments and thus we use a more systematic 
approach.}. 

In this work, we develop an improved fitting method by noticing
that the formulas for the baryon magnetic moments are linear
in the inertia parameters except $m{\cal I}_2$.
Thus, given a number for $m{\cal I}_2$,
a standard least-square-fitting method can be applied 
to determining the other five ${\cal M}$,
${\cal N}$, ${\cal N}^\prime$, ${\cal P}$ and ${\cal Q}$.
In Fig. 1, presented are the numerical results on the inertia 
parameters and the magnetic moments 
for various values of $m{\cal I}_2$ in the range
$0\le m{\cal I}_2\le 4$.
The fitting process turns out to be remarkably independent 
of the parameter $m{\cal I}_2$; the inertia parameters ${\cal M}$,
${\cal P}$ and ${\cal N}+\frac12{\cal N}^\prime$ are almost 
constants and the other parameters show a trivial 
linear dependence as 
\begin{equation}
\textstyle 
{\cal N} + \frac12{\cal N}^\prime = 10.95, \hskip 1em
m{\cal I}_2{\cal N}^\prime = 5.06 m{\cal I}_2 + 2.87, \hskip 1em  
{\rm and} \hskip 1em
{\cal Q} = 0.64 m{\cal I}_2 - 4.39.
\label{fit}\end{equation} 
Furthermore, for any value of $m{\cal I}_2$,
the formula given in Table 1 can
produce an excellent fit to the baryon magnetic moments
with almost constant $\chi^2$ of order $10^{-3}$
and $F^{(s)}_{2,p}$ remains at a {\em positive} constant 
value about +0.4. 
The numerical values are given in Table 2 as ``Fit1". 

Such an independence of the fitting to the parameter $m{\cal I}_2$
can be understood by using the V-spin symmetric relations
(\ref{v-spin}). By subtracting a baryon magnetic moment 
by that of the V-spin partner appearing in eq.(\ref{v-spin}), 
we can eliminate the terms with 
$m{\cal I}_2$ and ${\cal Q}$. 
It leads us to three equations 
for $\mu_p-\mu_{\Xi^-}$, $\mu_n - \mu_{\Sigma^-}$ 
and $\mu_{\Sigma^+} - \mu_{\Xi^0}$
for three unknowns ${\cal M}$, ${\cal N}+\frac12{\cal N}^\prime$
and ${\cal P}$. With the experimental values for the 
baryon magnetic moments the equations can be easily solved as 
${\cal M} = 2.51$, ${\cal N}+\frac12{\cal N}^\prime =10.76$
and ${\cal P} = -2.75$. 
These values are close to what we have 
obtained from the least-square fit including $\mu_\Lambda$.
Once these three parameters are fixed, 
the remaining equation can be used to determine 
the rest three inertia parameters. 
However, we can get only two equations for three unknowns,  
$m{\cal I}_2$, $m{\cal I}_2 {\cal N}^\prime$ and ${\cal Q}$,  
due to the last constraint in Eq.(\ref{v-spin}),  
$\delta \mu_{2, \Sigma^+} = \frac12 \delta \mu_{2,p}$.  
By using $\mu_p$ and $\mu_n$ we obtain 
$m{\cal I}_2{\cal N}^\prime = 4.96 m{\cal I}_2 + 3.34$, 
${\cal Q} = 0.64 m{\cal I}_2 - 4.91$, which are also comparable 
to Eq.(\ref{fit}).
It explains the linear dependence of parameters ${\cal Q}$
and $m{\cal I}_2{\cal N}^\prime$ on $m{\cal I}_2$. 
The constraint leaves a relation for the magnetic moments as
$ \mu_p + \mu_n - \mu_{\Sigma^-} + \frac12 \mu_{\Xi^-} 
- \frac32(\mu_{\Xi^0} + \mu_{\Sigma^+}) = 0$. 
With the experimental magnetic moments  the left 
hand side is evaluated as 0.08. 
It implies that the expressions 
given in Table 1 for the baryon magnetic moments are reasonable. 
The numerical results for this analysis are given as ``Fit2" in 
Table 2, which yields again a positive strange form factor. 

We present the numerical results of CBM\cite{cbm} also in Table 2 as a 
reference.  Here the inertia parameters are evaluated at the magic angle 
$\theta(R)=\frac12\pi$ corresponding to the bag radius $R\sim 0.6$fm. 
The values are obtained by including the conventional Skyrme term to the meson 
part of the Lagrangian in order to stabilize the soliton solution with the 
Skyrme parameter $e=4.75$ and $f_\pi = 93$ MeV.  The consequent baryon 
magnetic moments are presented in the next table, which show qualitative 
agreements with the experimental values within 30\% errors.  The worst case is 
the proton magnetic moment which comes out smaller than that of $\Sigma^+$.  
The same is true for the SM which corresponds to the CBM with zero bag radius.

Next we investigate the origin of the large positive values of the strange form
factors.  To do this we divide the nucleon strange form factors into three 
pieces; namely the contributions from the chiral symmetric limit
$F_{2N}^{(s),0}$, the explicit FSB via the current 
operator $F_{2N}^{(s),1}$ and the implicit FSB through
the representation mixing in the wave functions $F_{2N}^{(s),2}$ as shown in
Table 3.  Here one notes that the FSB effects are dominant in ``Fit1", ``Fit2" 
and the CBM through the explicit and implicit channels.  This large 
contribution from the FSB originates from the terms which represent the 
differences in the kaon and the pion decay constants $f_{K}\neq f_{\pi}$ and 
in the masses $m_{K}\neq m_{\pi}$ and $m_{s}\neq m_{u,d}$.  These terms 
implicitly describing the vacuum fluctuation in the strange direction have not 
been properly considered in most of models\footnote{Here one notes that one 
reference\cite{model2} treats the FSB effects to yield $F_{2,p}^{(s)}=-0.13$ 
n.m..  However this value has the same sign but is much larger than our SM 
prediction due to the fact that they used the different Skyrme parameter 
$e=4.0$ and missed in the inertia parameter $m$ the contribution from the term 
proportional to $f_{K}^{2}-f_{\pi}^{2}$.}.  That is why our prediction in 
``Fit1", ``Fit2" and the CBM are positively large differently from those of 
most of models. 

In Table 3 one also notes that the explicit current FSB is dominant in ``Fit1" 
and ``Fit2" while the implicit representation mixing FSB is assertive in the 
CBM.  Thus, one can hardly expect that those inertia parameters from the 
present state of chiral models could provide the ideal predictions of ``Fit1" 
and ``Fit2" on the magnetic moments and the strange form factors without an
introduction of new idea\footnote{Here one cannot exclude the possibility of
introducing new additional inertia parameters in the model independent 
relations which could yield a different prediction of the strange form factors, together with the gluon and Casimir effects.}.

The up-, down- and strange-quark contributions to the magnetic moment 
of the proton are presented in Table 2. 
The results show that the up-quark contribution is dominant 
to those from down- and strange-quarks by a factor 10. 
It is fully consistent with the up-quark dominant picture of Ref.\cite{mck}. 
However, this qualitative behavior could not be used to 
predict positive strange form factor. 
Note that the I-spin symmetry leads us to $\mu^{(u,d,s)}_p$ as 
\begin{equation} 
\textstyle
\mu^{(u)}_p = \frac23 [ 2\mu_p + \mu_n + F^{(s)}_{2,p}(0) ], \hskip 1em
\mu^{(d)}_p = \frac13 [ \mu_p + 2\mu_n + F^{(s)}_{2,p}(0) ], \hskip 1em
{\rm and} \hskip 1em
\mu^{(s)}_p = -\frac13 F^{(s)}_{2,p}.
\end{equation}
which would yield $\mu^{(u)}_p = 2.45$ n.m. and $\mu^{(d)}_p=-0.34$ n.m.  
in case of vanishing strange form factor.  Our theoretical prediction for the 
strange form factor of the other baryons is also given in Table 2.

In summary, we have investigated the magnetic moments and the strange 
form factors of the baryon octet based on the group structure of the chiral 
models.  Various symmetry breaking terms are included into the model 
and treated as perturbations in the SU(3) collective 
coordinate quantization scheme. 
Six inertia parameters appearing in the formulas for the magnetic 
moments are adjusted to fit the experimental values of the baryon octet.  
Our formulas are general to all the  
soliton models for the baryons with the SU(3) symmetry breaking 
terms treated as a perturbation.  We expect that in a more sophisticated 
version of CBM the situation is improved. 
The formulas turn out to fit the baryon magnetic moment remarkably well and 
predict the positive strange form factor of the proton 0.37 n.m. to be 
compared with the recent experimental result $0.23\pm 0.37\pm 0.15\pm 0.19$ 
n.m.\cite{sample}.  
We emphasize the role of the vacuum fluctuation in the strange 
direction to give the positive result, which is ignored in most of models.  

\vskip 1cm
We would like to thank Mannque Rho and G.E. Brown for helpful discussions and
constant concerns.  This work is supported in part by the Korea Science and
Engineering Foundation through the CTP and by the Korea Ministry of Education 
under Grant No. BSRI-97-2418.


\newpage
\begin{table}
\caption{Electromagnetic moments of baryon octet : expressions.}
\begin{center}
\begin{tabular}{ccccccc}
\hline
 $\mu_B$ & \hskip -0.5em=\hskip -0.5em & $\mu_{0,B}$ 
  & $\hskip -0.5em\oplus\hskip -0.5em$ & $\delta\mu_{1,B}$ 
  & $\hskip -0.5em\oplus\hskip -0.5em$ & $\delta\mu_{2,B}$ \\
\hline
$\mu_p$ 
 & \hskip -0.5em=\hskip -0.5em 
 & $+\frac{4}{40}{\cal M} + \frac{8}{30}({\cal N} + \frac12{\cal N}^\prime)$  
 & $\hskip -0.5em\oplus\hskip -0.5em$ 
 & $+\frac{16}{90}{\cal P} - \frac{8}{180}{\cal Q}$ 
 & $\hskip -0.5em\oplus\hskip -0.5em$
 & $m{\cal I}_2 [ \frac{24}{1500}{\cal M}
      + \frac{16}{2250}({\cal N} - 2{\cal N}^\prime)]$ \\
$\mu_n$
 & \hskip -0.5em=\hskip -0.5em 
 & $+\frac{2}{40}{\cal M} - \frac{6}{30}({\cal N} + \frac12{\cal N}^\prime)$
 & $\hskip -0.5em\oplus\hskip -0.5em$
 & $-\frac{10}{90}{\cal P} + \frac{14}{180}{\cal Q}$ 
 & $\hskip -0.5em\oplus\hskip -0.5em$
 & $m{\cal I}_2 [ \frac{62}{1500}{\cal M} 
      - \frac{92}{2250}({\cal N} - \frac{21}{23}{\cal N}^\prime ) ]$ \\
$\mu_\Lambda$
 & \hskip -0.5em=\hskip -0.5em 
 & $+\frac{1}{40}{\cal M} - \frac{3}{30}({\cal N} + \frac12{\cal N}^\prime)$
 & $\hskip -0.5em\oplus\hskip -0.5em$
 & $-\frac{9}{90}{\cal P} - \frac{9}{180}{\cal Q}$ 
 & $\hskip -0.5em\oplus\hskip -0.5em$
 & $m{\cal I}_2 [ \frac{27}{1500}{\cal M} 
      + \frac{18}{2250}({\cal N} - 2{\cal N}^\prime ) ]$ \\
$\mu_{\Xi^0}$
 & \hskip -0.5em=\hskip -0.5em 
 & $+\frac{2}{40}{\cal M} - \frac{6}{30}({\cal N} + \frac12{\cal N}^\prime)$
 & $\hskip -0.5em\oplus\hskip -0.5em$
 & $-\frac{22}{90}{\cal P} - \frac{4}{180}{\cal Q}$ 
 & $\hskip -0.5em\oplus\hskip -0.5em$
 & $m{\cal I}_2 [ \frac{12}{1500}{\cal M} 
      + \frac{8}{2250}({\cal N} - 2{\cal N}^\prime ) ]$ \\
$\mu_{\Xi^-}$
 & \hskip -0.5em=\hskip -0.5em 
 & $-\frac{6}{40}{\cal M} - \frac{2}{30}({\cal N} + \frac12{\cal N}^\prime)$
 & $\hskip -0.5em\oplus\hskip -0.5em$
 & $-\frac{8}{90}{\cal P} - \frac{8}{180}{\cal Q}$ 
 & $\hskip -0.5em\oplus\hskip -0.5em$
 & $m{\cal I}_2 [ \frac{24}{1500}{\cal M} 
      + \frac{16}{2250}({\cal N} - 2{\cal N}^\prime) ]$ \\
$\mu_{\Sigma^+}$
 & \hskip -0.5em=\hskip -0.5em 
 & $+\frac{4}{40}{\cal M} + \frac{8}{30}({\cal N} + \frac12{\cal N}^\prime)$
 & $\hskip -0.5em\oplus\hskip -0.5em$
 & $+\frac{26}{90}{\cal P} - \frac{4}{180}{\cal Q}$ 
 & $\hskip -0.5em\oplus\hskip -0.5em$
 & $m{\cal I}_2 ( \frac{12}{1500}{\cal M} 
      + \frac{8}{2250}({\cal N} - 2{\cal N}^\prime ) ]$ \\
$\mu_{\Sigma^-}$
 & \hskip -0.5em=\hskip -0.5em 
 & $-\frac{6}{40}{\cal M} - \frac{2}{30}({\cal N} + \frac12{\cal N}^\prime)$
 & $\hskip -0.5em\oplus\hskip -0.5em$
 & $-\frac{4}{90}{\cal P} + \frac{14}{180}{\cal Q}$ 
 & $\hskip -0.5em\oplus\hskip -0.5em$
 & $m{\cal I}_2 [ \frac{62}{1500}{\cal M} 
      - \frac{92}{2250}({\cal N} - \frac{21}{23}{\cal N}^\prime) ]$ \\
$\mu_{\Sigma^0}$
 & \hskip -0.5em=\hskip -0.5em 
 & $-\frac{1}{40}{\cal M} + \frac{3}{30}({\cal N} + \frac12{\cal N}^\prime)$
 & $\hskip -0.5em\oplus\hskip -0.5em$
 & $+\frac{11}{90}{\cal P} + \frac{5}{180}{\cal Q}$ 
 & $\hskip -0.5em\oplus\hskip -0.5em$
 & $m{\cal I}_2 [ \frac{37}{1500}{\cal M} 
		- \frac{42}{2250}({\cal N} 
			 - \frac{17}{21}{\cal N}^\prime) ]$ \\
\hline
$\mu_{N}^{(s)}$ 
 & $\hskip -0.5em=\hskip -0.5em$ 
 & $-\frac{7}{60}{\cal M} 
    + \frac{1}{45}({\cal N} + \frac12{\cal N}^\prime)$
 & $\hskip -0.5em\oplus\hskip -0.5em$
 & $+\frac{1}{45}{\cal P} + \frac{1}{90}{\cal Q}$
 & $\hskip -0.5em\oplus\hskip -0.5em$
 & $m{\cal I}_2 [ \frac{43}{2250}{\cal M}
		  - \frac{38}{3375}{\cal N}
		  + \frac{26}{3375}{\cal N}^{\prime} ]$ \\
$\mu_{\Lambda}^{(s)}$ 
 & $\hskip -0.5em=\hskip -0.5em$
 & $-\frac{3}{20}{\cal M}
   - \frac{1}{15} ({\cal N} + \frac12 {\cal N}^{\prime})$
 & $\hskip -0.5em\oplus\hskip -0.5em$
 & $-\frac{1}{15}{\cal P} - \frac{1}{30}{\cal Q}$
 & $\hskip -0.5em\oplus\hskip -0.5em$
 & $m{\cal I}_2 [ \frac{3}{250}{\cal M}
		  + \frac{2}{1125}{\cal N}
		  - \frac{4}{375}{\cal N}^{\prime} ]$ \\
$\mu_{\Xi}^{(s)}$
 & $\hskip -0.5em=\hskip -0.5em$
 & $-\frac{1}{5}{\cal M}
   - \frac{4}{45} ({\cal N} + \frac12 {\cal N}^{\prime} )$ 
 & $\hskip -0.5em\oplus\hskip -0.5em$
 & $-\frac{1}{9}{\cal P} - \frac{1}{45}{\cal Q}$
 & $\hskip -0.5em\oplus\hskip -0.5em$
 & $m{\cal I}_2 [ \frac{1}{125}{\cal M}
		   + \frac{4}{1125}{\cal N}
		   - \frac{8}{1125}{\cal N}^{\prime} ]$ \\
$\mu_{\Sigma}^{(s)}$
 & $\hskip -0.5em=\hskip -0.5em$
 & $-\frac{11}{60}{\cal M}
     +\frac{1}{15}({\cal N}+\frac{1}{2} {\cal N}^{\prime})$
 & $\hskip -0.5em\oplus\hskip -0.5em$
 & $+\frac{11}{135}{\cal P}+\frac{1}{54}{\cal Q}$
 & $\hskip -0.5em\oplus\hskip -0.5em$
 & $m{\cal I}_{2} [ \frac{37}{2250}{\cal M}
		  - \frac{14}{1125}{\cal N}
		  + \frac{34}{3375}{\cal N}^{\prime} ]$ \\
\hline
\end{tabular}
\end{center}
\end{table}

\clearpage
\renewcommand\arraystretch{1}
\begin{table}
\caption{The inertia parameters, magnetic moments of baryon octet 
and their form factors. The magnetic moments included in the fitting 
process are indicated by $*$. 
For the inertia parameters for ``Fit1", 
see the discussions in the text and Eq. (\ref{fit}).}
\begin{center}
\begin{tabular}{ccccccc}
\hline 
   & ${\cal M}$ & ${\cal N}$ & ${\cal N}^\prime$ & ${\cal P}$ 
   & ${\cal Q}$ & $m{\cal I}_2$ \\
\hline
Fit1 & 2.55 & text &  text   & $-2.91$ & text    & free \\
Fit2 & 2.51 & 7.73 & $+6.07$ & $-2.75$ & $-2.99$ & 3.00 \\
CBM  & 0.66 & 6.00 & $+0.52$ & $+1.11$ & $+1.27$ & 3.96 \\
SM   & 0.67 & 5.03 & $+0.91$ & $+0.76$ & $+0.99$ & 1.79 \\
\hline
\end{tabular}
\vskip 3ex
\begin{tabular}{cllllllll}
\hline
   & \multicolumn{1}{c}{$\mu_p$} 
   & \multicolumn{1}{c}{$\mu_n$} 
   & \multicolumn{1}{c}{$\mu_{\Lambda}$} 
   & \multicolumn{1}{c}{$\mu_{\Xi^0}$} 
   & \multicolumn{1}{c}{$\mu_{\Xi^-}$} 
   & \multicolumn{1}{c}{$\mu_{\Sigma^+}$} 
   & \multicolumn{1}{c}{$\mu_{\Sigma^0}$}
   & \multicolumn{1}{c}{$\mu_{\Sigma^-}$} \\
\hline
Fit1 & $2.80^*$ & $-1.91^*$ & $-0.58^*$ & $-1.28^*$ 
       & $-0.71^*$ & $2.40^*$ & 0.62 & $-1.16^*$ \\
Fit2 & $2.79^*$ & $-1.91^*$ & $-0.56^*$   & $-1.27^*$ 
       & $-0.69^*$ & $2.41^*$ & 0.62 & $-1.16^*$ \\
 CBM & 2.06 & $-2.03$ & $-0.58$ & $-1.43$ & $-0.49$ 
       & 2.12 & 0.43 & $-1.25$ \\
 SM  & 1.68 & $-1.33$ & $-0.59$ & $-1.24$ & $-0.52$ 
       & 1.76 & 0.54 & $-0.68$ \\
\hline
Exp. & 2.79 & $-1.91$ & $-0.61$ & $-1.25$ 
       & $-0.69$ & 2.43 & $-$  & $-1.16$\\
\hline
\end{tabular}

\vskip 3ex
\begin{tabular}{crrrrrrrr}
\hline
$ $ & \multicolumn{1}{c}{$\mu^{(u)}_p$} 
    & \multicolumn{1}{c}{$\mu^{(d)}_p$} 
    & \multicolumn{1}{c}{$\mu^{(s)}_p$} 
    & \multicolumn{1}{c}{$F^{(s)}_{2N}$}
    & \multicolumn{1}{c}{$F^{(s)}_{2\Lambda}$ }
    & \multicolumn{1}{c}{$F^{(s)}_{2\Xi}$}
    & \multicolumn{1}{c}{$F^{(s)}_{2\Sigma}$}
    & \multicolumn{1}{c}{$F^{0}_{2}$}\\
\hline
Fit1  & 2.72 & 0.21 & $-0.13$ &   0.39  & 1.43 & 1.26 & $-0.97$ & 0.28    \\
Fit2  & 2.70 & 0.22 & $-0.12$ &   0.37  & 1.37 & 1.22 & $-0.99$ & 0.26    \\
 CBM  & 1.59 & 0.57 & $-0.10$ &   0.30  & 0.49 & 0.25 & $-1.54$ & $-0.67$ \\
  SM  & 1.34 & 0.33 & $+0.01$ & $-0.02$ & 0.51 & 0.09 & $-1.74$ & $-0.67$ \\
\hline
\end{tabular}
\end{center}
\end{table}
\begin{table}[t]
\caption{The strange form factors of nucleon.}
\begin{center}
\begin{tabular}{crrrr}
\hline
$ $  &$F_{2N}^{(s),0}$ &$\delta F_{2N}^{(s),1}$ &$\delta F_{2N}^{(s),2}$ 
     &$F_{2N}^{(s)}$\\
\hline
Fit1 &0.16 &0.28    &$-0.05$ &0.39\\
Fit2 &0.16 &0.28    &$-0.07$ &0.37\\
CBM  &$-0.19$ &$-0.12$ &0.61 &0.30\\
SM   &$-0.13$ &$-0.09$ &0.20 &$-0.02$\\
\hline
\end{tabular}
\end{center}
\end{table}
\newpage
\begin{figure}
\caption{Numerical results of the fitting : (a) inertia parameters, 
(b) $\chi^2$, (c) baryon magnetic moments and (d) proton strange 
form factor as a function of $m{\cal I}_2$.  The solid lines in (c) are the 
experimental values of the baryon magnetic moments.}
\vskip 4ex
\begin{center}
\epsfig{file=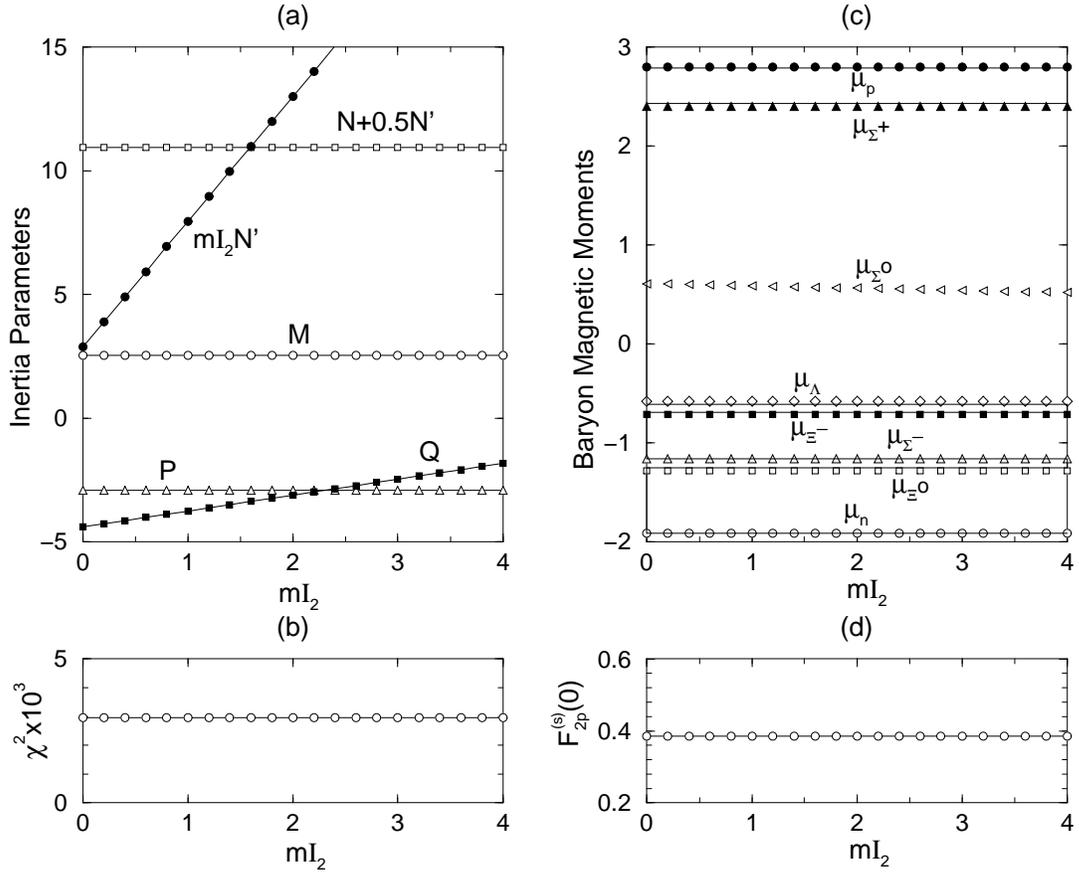, width=15cm}
\end{center}
\end{figure}
\end{document}